\documentclass[12pt,preprint]{aastex}
\usepackage{psfig}
\usepackage{graphicx}
\def\msun{\mbox{M$_{\odot}$}}
\shorttitle{Helium Stars}
\shortauthors{authors}

\begin{document}

\title{The Role of Helium Stars in the Formation of Double Neutron Stars}

\author{N.\ Ivanova\altaffilmark{1}, K.\ Belczynski\altaffilmark{1,2}, V.\ 
Kalogera\altaffilmark{1}, F.A.\ Rasio\altaffilmark{1}, \& R.E.\ 
Taam\altaffilmark{1}}

\affil{ $^{1}$ Northwestern University, Dept. of Physics \& Astronomy,
       2145 Sheridan Rd., Evanston, IL 60208\\
        $^{2}$  Lindheimer Postdoctoral Fellow\\
nata, belczynski, vicky, rasio, r-taam@northwestern.edu}

\begin{abstract}{
We have calculated the evolution of 60 model 
binary systems consisting of helium stars in the 
mass range of $M_{\rm He}= 2.5 - 6\,\msun$ 
with a 1.4 $\msun$ neutron star companion 
to investigate the formation of double neutron star systems.  
Orbital periods ranging from 0.09 to 2 days are considered,
corresponding to Roche lobe overflow starting  
from the helium main sequence to  
after the ignition of carbon burning in the core. 
We have also examined the evolution into a common envelope 
phase via secular instability, 
delayed dynamical instability, and the consequence of matter filling the  
neutron star's Roche lobe. 
The survival of some close He-star neutron-star binaries through the last
mass transfer episode 
(either dynamically stable or unstable mass transfer phase) 
leads to the formation of extremely short-period double neutron star 
systems (with $P\lesssim 0.1$
days). In addition, we find that 
systems throughout the entire calculated mass range can evolve into a common 
envelope phase, depending on the orbital period 
at the onset of mass transfer.  
The critical orbital period below which  common envelope 
evolution occurs generally increases  with $M_{\rm He}$. 
In addition, a common envelope phase may occur during a
short time for systems characterized by orbital periods
of 0.1 - 0.5 days at low He-star masses 
($\sim 2.6 - 3.3\,\msun$).  

The existence of a short-period population of double neutron stars increases   
the predicted detection rate of inspiral events by 
ground-based gravitational-wave detectors and impacts their  
merger location in host galaxies and their possible role as $\gamma$-ray burst 
progenitors. We use a set of population synthesis calculations and
investigate the implications of the mass-transfer results for
the orbital properties of DNS populations.
} 
\end{abstract}


\section{INTRODUCTION}  

The formation of double neutron star (DNS) systems is thought to be the
endpoint of long sequences of evolutionary stages involving massive binaries
(e.g., Bhattacharya \& van den Heuvel 1991, Belczynski et al. 2002b).
Although a complete understanding of the detailed DNS
formation channels remains to be attained, the observational characteristics 
of observed DNS systems clearly indicate the involvement of massive stars and
mass-transfer  phases that can lead to dramatic orbital contraction (e.g.,
common-envelope phases) prior to the formation of the second-born neutron
stars.  Uncovering the details of DNS formation and its dependence on the
outcomes of prior binary evolution phases has important implications for our
understanding of binary-pulsar formation (van den Heuvel \& Taam 1984),
supernovae of hydrogen-poor stars (Nomoto et al. 
1994), sources of gravitational waves 
(Clark \& Eardley 1977), and possibly $\gamma$-ray bursts (Paczynski 1986).

One of the intermediate evolutionary stages of DNS 
formation involves binaries  
with first-born neutron stars (NS) and helium-rich companions. This stage
follows that of the high-mass X-ray binaries (NS with massive, 
hydrogen-rich
companions). In this study we focus on the role of such helium-rich NS
companions in DNS formation mechanisms. In particular we investigate the
response of helium stars to binary mass transfer driven by Roche lobe
overflow. Our calculations are in some ways complementary to those
presented by Dewi et al. (2002), and we address the issue of the outcome of
mass transfer episodes in the context of DNS formation. We are especially
interested in (i) the degree of orbital contraction (if any) during mass
transfer driven by the helium star, (ii) the possible development of 
mass transfer on dynamical time scales and of a common-envelope (CE) phase
\cite{TSreview}, and (iii) the possible merger of the binary, 
aborting DNS formation.

The outcome of this last episode of mass transfer in DNS progenitors has
important implications for the distribution of orbital separations and,
hence, of gravitational-wave merger lifetimes of DNS systems and the 
location of merger sites relative to host galaxies. In this context, 
Belczynski et al. (2002b) 
have explored the implications of the {\em assumption} 
that {\em low-mass} helium 
giants (M $\lesssim 4.5 \msun$) \cite{Habets, WLW95} develop sufficiently deep 
convective envelopes to drive mass transfer onto NS on dynamical
time scales and therefore evolve into a CE phase. 
Under this assumption
they found that a large fraction of binaries can survive this phase.  
Subsequent evolution leads to the explosion of the helium star core in a Type 
Ic supernova 
and the formation of very tight DNS systems with merger lifetimes shorter
than a few Myr. Systems with such short lifetimes would be difficult to detect
as binary radio pulsars relative to the known systems with longer lifetimes
\cite{BKB} and would not have sufficient time to escape their host
galaxies \cite{BBK, BBR, PB} before merging.  
We point out that Belczynski et al. (2002b) 
had also taken into account the effects of 
hyper-critical accretion onto the NS during the CE phases \cite{Brown}. 
Despite some accretion, both during the H-rich CE phase (end of X-ray binary
stage) and the He-rich CE phase (last binary interaction) in DNS formation
scenarios, the NS avoids collapse into a black hole.
These results are based on calculations of the mass accretion rate onto NS and the  
assumption that the maximum NS mass is greater than $\simeq 
1.5 \msun$ [for more details see Belczynski et al. (2002a)].

In this study we perform detailed binary evolutionary calculations relevant 
to helium stars overfilling their Roche lobes and transferring matter to 
their NS companions. The calculations cover a range of helium-star masses 
($M_{\rm He} = 2.5 - 6\, \msun$) and evolutionary stages that are relevant to DNS
formation. For each  sequence, the temporal evolution of the mass-transfer
rate and orbital  period are calculated and the outcome of the mass transfer (dynamical
or thermal  instability or merger) is examined. Since thermal instability can
lead to high mass transfer rates, in excess of the NS Eddington limit, we
examine its consequences in the context of considerations of the {\em
trapping radius} [i.e., the characteristic radius for mass loss due to
super-Eddington mass transfer \cite{Beg79}] and the possibility that a 
common-envelope phase develops \cite{KBeg}. 
The results delineate the parameter regime (mass of the helium star and 
the orbital period at the onset 
of mass transfer) where systems survive as binaries from those where the 
systems merge. 
Using the {\em StarTrack} population synthesis code \cite{BKB} we examine 
the implications of these results for the physical
characteristics of DNS populations and present our results.

In the following section we describe our computational methods and initial
models both for the mass transfer sequences and the DNS population synthesis
calculations. In section \S\,3  we describe our results in detail and  
in section \S\,4 we discuss their implications for DNS formation, 
binary models for
hydrogen-poor supernovae, gravitational waves, and $\gamma$-ray bursts.

\section{METHODS AND INITIAL MODELS}
\subsection{Mass-Transfer Sequences} 

For the mass transfer calculations we have used 
a standard Henyey-type stellar evolution code 
\cite{KWH}, recently updated by Podsiadlowski et al. (2002). 
We have adopted OPAL opacities \cite{RI}, 
supplemented with contributions from atomic, 
molecular and grain absorption in the low temperature regime ($\lesssim 12,500$
K) \cite{AF}.  The nuclear reactions (tracking 40 isotopes) describe the 
major burning stages 
through oxygen burning with rates taken from
Thielemann's library REACLIB  \cite{TTA} and updated 
as in Cannon (1993)\footnote{The reaction rates have been compared to 
the recent compiled NACRE reaction rates (Angulo et al 1999). 
Differences of less than 5\% are found for important reaction rates
in the required temperature range. Since the triple-$\alpha$ reaction
rates are highly temperature sensitive the differences are 
inconsequential for the evolution.}.

We have calculated models for zero-age Main-Sequence helium stars (HeMS) 
with a solar metallicity ($Y=0.98, Z=0.02$). Since such stars are characterized by 
convective cores and can evolve to include convective envelopes, we have adopted the
Schwarzschild convection criterion with a mixing-length 
parameter $\alpha = 2$ and a convective-overshooting parameter 
equal to 12\% of the pressure scale height $H_P$. 
We have also included in the models mass loss associated with stellar winds 
with rates given by Hurley et al. (2000) as 
\begin{equation}  \dot M_{\rm He, wind} = - \max \left[2\times 10^{-13} {\frac
{LR} {M}}, 10^{-13}L^{1.5}\right] {\rm M}_\odot\,{\rm yr}^{-1} \ . 
\end{equation}  Here, M, R, and L are the stellar mass, 
radius and luminosity in solar units.   

To determine the adequacy of our stellar models, we have carried out
evolutionary sequences to 
the stage of oxygen core burning for single helium stars 
with masses in the range of 2 to 6 $\msun$.   
We have found the internal structure as well as the radii of our models at
various evolutionary stages to be in good agreement with
results obtained by Habets (1986), Dewi et al. (2002),
Pols (2002), and Woosley (1997).

We have calculated binary mass-transfer sequences assuming a 
NS mass $M_{\rm NS} = 1.4 {\rm M}_\odot$ and initial helium star masses 
$M_{\rm He; 0}$ in the range 2.5--6\,M$_\odot$. 
We have chosen the orbital
period $P_{\rm tr}$ at the onset of the mass transfer as the main input parameter
for our mass transfer sequences. 
For a given value of $P_{\rm tr}$ and $M_{\rm He; 0}$ 
we calculated the age at which this star 
will overfill its Roche lobe by evolving the helium star from the ZAMS
to the onset of mass transfer as a single star.
We typically start the 
binary sequences at an evolutionary age slightly before 
Roche lobe overflow 
and followed the evolution until core-oxygen burning, unless 
we encounter conditions leading to a binary merger.

During Roche lobe overflow, mass transfer
rates are  calculated following the prescription adopted in 
Tout \& Eggleton (1988)
 \begin{equation}
 \dot M_{\rm tr} = 10^3 \times \max \left[0, (\ln(R/R_{\rm RL })^3      
\right]\,{\rm }M_\odot\,{\rm yr}^{-1}, 
 \end{equation}
 where $R_{\rm RL }$ is the Roche-lobe radius and is 
given as a function of the binary mass ratio $q\equiv M_{\rm He}/M_{\rm NS}$ 
and the orbital separation $A$ by \cite{Eg83}
 \begin{equation}
 R_{\rm RL } = {\frac {0.49 q^{2/3} } {0.6 q^{2/3}+\ln (1+q^{1/3}) } } A\ .
 \end{equation}
During detached phases we also account for the 
effects of mass transfer due to {\em atmospheric} Roche lobe overflow. We use a 
prescription by Ritter (1988)
for the mass transfer rate calculation
 \begin{equation}
 \dot M_{\rm tr} = \dot M_0 \exp \left ({\frac {R_{\rm RL} -R}
{H_P}}\right)\ , 
 \end{equation}
 where $\dot M_0$ is the mass transfer  rate when the 
mass losing star fills
its Roche lobe, $\dot M_0={\frac {1} {\sqrt{e}}} \rho_{\rm ph} v_s Q$.
Here  $\rho_{\rm ph}$ is the photospheric density of the helium star,  
$v_s$ is the isothermal sound speed and 
$Q\approx  R_{\rm RL } H_{\rm p}$ 
is the effective cross section of the gas stream at the inner Lagrangian 
point, L$_1$.
The total rate at which the helium star loses mass is  
\begin{equation}
\dot M_{\rm He} = -\dot M_{\rm tr} + \dot M_{\rm He, wind}.
\end{equation}

Throughout our mass-transfer sequences we have assumed 
that NS can accrete only through the mass transfer process, i.e.,
there is no wind accretion and the mass accretion rate
is Eddington limited: 
 \begin{equation}
\dot M_{\rm  accr} = \min\{\dot M_{\rm tr}, \dot M_{\rm  edd}\}\ ,
 \end{equation}
where
 \begin{equation}
 \dot M_{\rm  edd} \equiv 
{\frac {4 \pi c R_{\rm NS}} {\kappa} } \ . 
 \end{equation}
Here $c$ is the speed of light, 
$R_{\rm NS}$ is the radius of the NS (assumed to be 10 km)
and $\kappa$ is the electron scattering opacity of helium rich material.
The orbital evolution  
is calculated assuming that, if the mass transfer rate 
exceeds the NS Eddington 
limit, the excess material escapes 
the system isotropically with the NS specific orbital    
angular momentum. This mass loss combined with the 
isotropic wind mass loss leads to loss of binary orbital angular
momentum at a rate given by \cite{vdH94, KW96, SPvdH97} 
 \begin{equation}
 {\frac {\dot J_{\rm ML}} {J_{\rm orb}}} = {\frac {\gamma + \beta q^2} {1+q}}
{\frac {\dot M_{\rm He}}{M_{\rm He}} }, \
 \end{equation}
 where $\gamma \equiv \dot M_{\rm He, wind} / \dot M_{\rm He} $ and $\beta
\equiv - (\dot M_{\rm tr} - \dot M_{\rm accr} )/ \dot M_{\rm He}$. 

We have also included angular momentum losses 
due to gravitational radiation at a rate (Landau \& Lifshitz, 1958)
 \begin{equation}
 {\frac {\dot J_{\rm gw}} {J_{\rm orb}}} = -{\frac {32 G^3} {5 c^5}}
{\frac {M_{\rm He} M_{\rm NS} ({M_{\rm He} + M_{\rm NS}})} {A^4} } \ .
 \end{equation}

We have examine 60 models for helium stars with  
masses in the range 2.5--6\,$\msun$ for a wide range of orbital separations. 
Although we mainly focus on sequences where mass transfer is initiated after 
the end 
of core-helium burning, we also include a few calculations for systems that 
reached Roche lobe overflow (RLOF) during the core helium burning phase. 
The initial parameters  
for the representative mass-transfer sequences are given in Table 1.

\subsection{Population Synthesis Models}

To examine the effects of the calculated mass-transfer sequences on  
the physical properties of DNS populations, we have incorporated
our results 
into a recently developed population synthesis code {\em StarTrack}. A detailed
description of this code can be found in  Belczynski et al. (2002b). 
With StarTrack we can follow the evolution of a large ensemble of single
and binary stars through long evolutionary sequences, for a large range of
masses and metallicities.  The single star evolution is based on the
analytic fits provided by Hurley et al. (2000) with some modifications
related to the determination of compact object masses. The modeling of
binary evolution incorporates detailed treatment of stable and unstable,
conservative and non-conservative mass transfer  episodes, mass and angular momentum
loss through stellar winds (dependent on metallicity) and gravitational
radiation, hyper-accretion onto compact objects, and asymmetric core
collapse events with a realistic spectrum of compact object masses.  It
further includes an orbit calculator that allows us to model the motion of
all systems in gravitational potentials of galaxies
(Belczynski et al. 2002c).

Common-envelope evolution is calculated based on the commonly used
energy-balance formulation (Tutukov \&
Yungelson 1979), where orbital energy is
consumed with a certain efficiency $\alpha_{\rm CE}$ to balance the envelope
binding energy $\alpha_{\rm CE}\Delta E_{\rm orb}=E_{\rm bind}$. 
The latter energy is proportional to the inverse of a numerical
factor $\lambda$ that describes the degree of central concentration of the
mass donor $E_{\rm bind}\propto 1/\lambda$ 
(de Kool 1990; Dewi \& Tauris 2000). 
The product $\alpha_{\rm CE}\lambda$ is typically treated 
as a free parameter in the models, and in the present
study we explore 3 models: 1, 0.3 and 3.
We note that although $\alpha_{\rm CE}$ and $\lambda$ are 
physically separate parameters,
varying them independently is mathematically equivalent to
varying their product.
The NS masses are obtained from a mapping between the final CO 
core mass of a given pre-SN star (Hurley et al.
2000) and the final Fe-Ni core mass for a given CO core mass based on 
models of Woosley (1986).  Since 
hydrodynamical calculations (Fryer 1999) show that there is little  
fall back during NS formation, 
we assume that the Fe-Ni core collapses and forms a NS, with a mass  
equal to the baryonic mass of the collapsing core.
The rest of the pre-SN star is lost in the explosion. 
Asymmetric supernova events are modeled with NS natal kicks of random
direction and magnitudes drawn from the Cordes \& Chernoff (1998)
distribution (a weighted sum of two Maxwellians, one with $\sigma=175$\    
km~s$^{-1}$ (80\%) and the second with $\sigma=700$~km~s$^{-1}$ (20\%).
In this study all other model parameters are assumed as in Model A in 
Belczynski et al. (2002b). 

As noted in the Introduction, the version of StarTrack used in Belczynski
et al. 2002b adopted the following set of assumptions regarding the
behavior of Roche-lobe filling helium stars: (i) helium stars that
reach Roche overflow during HeMS were assumed to lead to mergers;
(ii) low-mass ($\le 4\, M_\odot$) 
helium stars that have evolved beyond the HeMS
were assumed to have developed sufficiently massive convective envelopes
to drive mass transfer on dynamical time scales leading the binary to a CE
phase. Evolution through the CE phase was treated as described above.
In the present study we compare the properties of DNS populations formed
under these assumptions and under two other sets 
of assumptions that are developed
based on our detailed  mass transfer calculations presented here. 
These two new sets of assumptions are described in \S\,3.2.

\section{NUMERICAL  RESULTS}

\subsection{Mass-Transfer Sequences} 

Examination of the 60 model sequences reveals that 
He-star NS binaries can experience stable or unstable  
mass transfer episodes.  
Systems which 
enter the mass-transfer phase during or after He-shell burning are found 
to exhibit the
same qualitative behavior. For example, the evolution of a
binary with a 4\,$M_\odot$ He-star filling its Roche lobe during the He-shell
burning stage (initial orbital period equal to 0.1\,d) is illustrated in 
Figure 1. It can be seen that the 
binary experiences two episodes of mass transfer, first 
during He-shell burning and the second
after core-C ignition.  This result is typical for all such close binary 
systems in which the mass transfer  is
initiated during He-shell burning and is in agreement with the results
obtained by Dewi et al. (2002). Between these two mass transfer  episodes the system is 
detached for a few thousand years. 
During this detached stage the helium star is
entirely radiative; the entropy in the very outer layers of the 
helium star envelope increases.
In particular, for periods $P_{\rm tr}\ge 0.2$\,d ($M_{\rm He;0}\le3.5$\,$M_\odot$) and 
$P_{\rm tr}\ge 0.3$\,d ($M_{\rm He;0}>3.5$\,$M_\odot$),
the entropy in the outer radiative layer 
increases sufficiently that the mass transfer rate during the
second mass transfer event is generally higher than during the first stage. 
On the  other hand, for a binary that entered Roche lobe overflow shortly before 
the core-C ignition stage, mass transfer is uninterrupted
(e.g., for binaries with orbital period at the onset 
of mass transfer of $P_{\rm tr}\simeq 0.3$\,d).
Similarly, binaries with more evolved helium stars (past He-shell
burning) and wider orbital separations experience only one mass transfer  episode after 
core-C ignition.

The binary separation decreases throughout 
the evolution as a result of the mass ejection from the system.  For the 
example illustrated in Figure~1, the orbital period decreased by about a 
factor of 2.
In general, for the majority of sequences, 
the orbital period is decreased by factors as large as 4. 
We note that, for the system presented in Figure 1, 
the mass transfer rate exceeds
the NS Eddington accretion rate during the entire Roche-lobe overflow phase, 
and the bulk of the He-rich material is ejected from the binary with the 
specific orbital angular momentum of the NS. 

For those systems in which mass transfer becomes unstable, the systems may 
enter a CE phase.  It is well known that there 
are two circumstances under which a binary system can undergo CE evolution 
\cite{TSreview}: either due to a secular instability or due to a mass transfer  episode on 
a dynamical timescale. 

For the secular instability to develop, the moment of inertia of the non-compact 
and larger star must exceed one third of the orbital moment of 
inertia of the binary system (e.g., Hut 1980, for a review see Rasio 1996).
If the moment of the inertia of the helium star is written as
$I_{\rm He} = k^2 M_{\rm He}R^2_{\rm He}$, where $k$ is the dimensionless
gyration radius of the He-star,
the condition for the development of the instability is
 \begin{equation}
 k^2 > {\frac {1} {3 f^2 r^2_{\rm RL}}} {\frac {1}{1+q}} \ ,
 \end{equation}
 where $f\equiv R_{\rm He}/R_{\rm RL}\le 1$, and $r_{\rm RL} = R_{\rm RL}/A$. 
This condition leads  
to $k^2>0.25/f^2$ for $M_{\rm He; 0}=6 M_\odot$ and $k^2>0.65/f^2$ for
$M_{\rm He; 0}=2.5 M_\odot$. Note that 
$k^2=0.4$ for a uniform solid sphere. Since the value of $k$ decreases 
with more evolved phases and 
$k^2 = 0.08$ for a zero age HeMS, this condition can never be satisfied for He
stars, and the systems with He-star and NS components are secularly stable.

In contrast, mass transfer can be driven on a dynamical timescale, if the  
helium star develops a sufficiently deep and massive convective envelope.
In this case, the envelope tends to expand on a dynamical timescale
upon mass loss \cite{Web85}. It was shown by Habets (1986) 
that only helium stars 
less massive than 2.5\,$M_\odot$ would develop deep and massive
convective envelopes. More massive evolved helium stars can also develop 
substantial convective envelopes characterized by a large {\em radial} extent 
in the late evolutionary stages. 
However, for these stars only a very small mass fraction may be contained in the 
outer convective zone. 
We find that single helium stars 
of mass greater than 3\,$M_\odot$, evolved with 
wind mass loss, do not develop a sufficiently massive outer convective 
zone to initiate a CE phase since the mass contained in this zone is smaller than 
$0.01\%\,M_{\rm He}$.
To illustrate this point, we show in Figure 2 (top panels) the internal 
structure of a 4\,M$_\odot$ He-star just after it has evolved off the MS 
and after it has begun core-C burning. 
The convective zones are shown as hatched regions on the entropy plots.
For this model we found convection only in the center of the He-star.
Had we evolved a 4\,M$_\odot$ model without mass loss (not realistic, 
as He-stars 
are known to have significant wind mass loss) we would have found a small 
outer convection zone ($\lesssim 0.03\,M_\odot$).
In both cases (either no or very little mass in the outer convection zone) the
He-star behaves as a radiative star contracting upon mass loss \cite{Web85}.  
We note that, although the evolved helium stars (after core He burning) more 
massive than $\sim 3\, M_\odot$ transfer mass at highly 
super-Eddington rates 
(the Eddington limit for helium accretion $\dot M_{\rm 
NS, edd}\approx 3\times 10^{-8}\,\rm M_\odot {\rm yr}^{-1}$), 
the rates are not 
high enough to drive dynamically unstable mass transfer  and initiate 
a CE phase (e.g., see Figure 1). 
Only for He-stars less massive than 2.5\,M$_\odot$ and for most evolved He-stars 
(after core-C ignition) in binaries with orbital periods at the onset of mass transfer  in 
excess of $\sim 1$  year 
does a substantial convective envelope develop, leading to a 
dynamical mass transfer instability and CE evolution.

For donor stars with mostly radiative envelopes, it has been pointed
out that a {\em delayed} dynamical instability can develop
\cite{Web85, Hjellming}. This occurs during the thermally unstable 
mass transfer phase (where the mass ratio of donor to accretor is greater 
than 1.4) when the steep surface entropy profile has been removed.  
For a donor star that has not had sufficient time to relax to thermal 
equilibrium, deeper layers with flat entropy profile can be exposed.  
Subsequent mass loss then leads then to stellar expansion. 
Typical critical mass ratios above which the delayed dynamical instability 
occurs lie between 2, at the beginning of the hydrogen main
sequence, and 4, near the base of hydrogen
giant branch \cite{Hjellming}.  In the present 
calculations we find that the development of the delayed dynamical
instability is unavoidable in binaries in which mass transfer  is initiated on or
soon after the HeMS for the entire studied He-star 
mass range and for all He-stars 
(independent of their evolutionary stage at the onset of mass transfer) in binaries with 
mass ratios exceeding 4.
The question then arises as to whether these binaries can
survive the ensuing CE phase. Our calculations suggest 
that they are not likely to survive. 
In the lower panels of Figure~2 we show the orbital energy 
as well as the minimum energy $E_{\rm min}$ needed to unbind 
envelope material for two evolutionary stages.
The quantity $E_{\rm min}$ is determined from an 
integration over the mass contained above the core to the stellar surface 
as 
\begin{equation}
E_{\rm min} =
\int_{M_{\rm core}}^{M_{\rm He}}  \left (-{\frac  {GM_r} {r}} + H(r)\right) 
dm\ ,
\end{equation}
where $H(r)$ is the {\em specific enthalpy\/}, 
$M_{\rm  core}$ is the mass of the core of helium star, 
and $M_r$ is the mass inside a given radius $r$. 
The internal energy that is included in the enthalpy 
corresponds to that of a fully ionized gas decreased by the ionization energy 
in partially ionized layers of the envelope. We assume that 
the recombination energy does not play a role in driving the helium 
envelope expansion. 
The change in the orbital energy during the spiral-in process   
is given by 
\begin{equation}
\Delta E_{\rm orb} = 
 {\frac {GM_{\rm  core}M_{\rm NS}}{2A_f}} - {\frac {GM_{\rm HE}M_{\rm NS}}{2A}}
 \ ,
\end{equation}
where $A_f$ is the final (after the CE-phase) orbital separation.
Soon after the HeMS the orbital energy is smaller than the envelope binding 
energy except very deep in the He-star interior (see Figure 2). 
Examination of models at masses other than $4 \msun$ show the same
quantitative behavior. Therefore, we conclude that the
outcome of the mass transfer  and following CE phase in such close binaries is a merger. 
If mass transfer  and the delayed dynamical instability develop during or after core-C
burning, the binary might avoid a merger, but only if the transfer of energy 
from the orbit to the envelope is efficient ($\gtrsim 50$\%; see 
Figure 2).
If only about half or less energy will be converted to the energy driving 
the expansion of the envelope, the final binary separation will have been 
reduced to such an extent that the core of the helium star overfills
its own Roche lobe and a merger will take place. 
However, if a significant fraction of the envelope ($> 50\%$) is removed 
prior to the common envelope phase, then 
the binary could survive the CE.

In addition to the above two channels for CE evolution, 
systems involving mass transfer onto a NS may also develop  
a CE provided that the trapping radius exceeds 
the Roche lobe radius of the NS \cite{KBeg}. 
The trapping radius is the radius at which the luminosity
generated by infall of the material reaches the Eddington limit.
It is calculated by setting the infall speed of the gas equal
to the approximate outward photon diffusion speed of the 
radiation (for details see King \& Begelman 1999).
Interior to this radius photon diffusion outwards cannot overcome 
the advection of matter moving inwards. 
If the accreting compact object is a black hole,
the radiation that is generated in excess of the
Eddington limit can be swept into the black hole and lost.
However, if the compact object is a neutron star, radiation
pressure resists inflow at a rate in excess of the Eddington limit,
causing the stellar envelope to grow outwards.

Here the trapping radius, where 
expulsion of some of the material transferred from a 
star takes place, is given by \cite{Beg79}
 \begin{equation}
 R_{\rm trap} = {\frac {\dot M} {\dot M_{\rm NS, edd}}} 
{\frac {R_{\rm NS}} {2}} \ ,
 \end{equation}
where $\dot M$ is the accretion rate. For the case when the accreted 
matter is helium rich one finds
 \begin{equation}
 R_{\rm trap} =  6.4\times 10^{13}\ {\rm [cm]}\ \dot m\ \ \ ,   
 \end{equation}
where $\dot m$ is the accretion rate in M$_\odot$\,yr$^{-1}$.
This criterion can be reexpressed 
in terms of a critical mass transfer  rate.  If the mass transfer  rate driven by the donor
is higher than the rate with which the NS can expel the material (using the
Eddington accretion luminosity as powering mechanism), then a CE may form.
Using the fact that the NS is the less massive component of the 
system, its Roche lobe
in solar radii is $R_{\rm rl} = 1.9 M^{1/3}_{\rm  NS}P^{2/3}_{\rm d}$ 
(here $P_{\rm d}$ is the orbital period in days).
The estimate for the critical mass transfer rate is then:
 \begin{equation}
 \dot M_{\rm crit} \approx 2\times 10^{-3} M_{\rm NS}^{1/2} P_d^{2/3}
\ {\rm M}_\odot\,{\rm yr}^{-1}.  
\label{mcrit}
 \end{equation}
We estimate that systems with the maximum mass transfer  rate of about 
$\dot M_{\rm trap} \approx 10^{-3}$ 
$M_\odot$ yr$^{-1}$ and orbital periods $P_d \le 0.3^d$ 
will satisfy the condition that 
$\dot M_{\rm trap} > \dot M_{\rm crit}$ for CE formation.

Although our current understanding is not sufficiently developed to conclude 
with certainty that CE evolution will ensue if the above critical rate is 
exceeded, this remains a possibility. In this context, comparison of 
the calculated mass transfer  rates of our sequences with $\dot M_{\rm crit}$ leads to the 
conclusion that the formation of a CE via this mechanism is, indeed, 
possible (see Figure 1).  The duration of this unstable phase (when the mass transfer  
rate exceeds
$\dot M_{\rm crit}$) depends on the evolutionary stage of the helium star, and ranges 
from a few hundred years, if the critical mass transfer rate is achieved 
during He-shell burning (case B), to a
few years, if the critical mass transfer rate is achieved after 
core-C ignition (case B or C). 
In the latter case
the spiral-in timescale could be 
longer than the  helium star evolutionary timescale. This is expected since,
in this case, the mass in the stellar envelope is comparable to 
or less than the NS mass
and significant spin-up of the common envelope is possible, prolonging
the spiral-in phase (Sandquist et al. 2000).  
We therefore expect that for a  
CE phase developing in the very advanced evolutionary 
phases of the helium star the spiral-in will
not be completed before the helium star ends its 
evolution in a supernova and a DNS forms.

The overall results of the most important evolutionary sequences are 
listed in Table 1.  Specifically, the final orbital periods $P_{\rm f}$, total 
mass $M_{\rm He; f}$ and envelope mass $M_{\rm env; f}$  of the evolved helium stars are presented.  
The mass of the helium envelope is given at the end of the first mass transfer  stage 
as well as the final envelope mass at the end of the evolution.   
The mass of the helium envelope is defined as the mass 
above the radius where the helium abundance is less than $1\%$. 

In addition, 
we also provide an estimate of the thermal mass transfer rate, 
$\dot M_{\rm TH}$,
(i.e., the maximum mass transfer rate for which the star can remain in 
thermal equilibrium), where 
$\dot M_{\rm TH} = {\Delta M_{\rm env}\over t_{\rm TH}}$
and 
$t_{\rm TH} = 1.5 \times 10^7 {M_{\rm He}^2\over {R_{\rm He} L}}$ yr.
Here $R_{\rm He}$ and L are the radius and luminosity of the helium star in 
solar units.  It can be seen from Table 1 that the average  
mass transfer  rate (for Case B) is greater than $\dot M_{\rm TH}$ in sequences 
for which the mass transfer  is initiated at an earlier stage.  
Here, the average mass transfer  rate is given by the ratio of the mass lost to the 
duration of the mass transfer  phase.
On the other hand, 
mass transfer  initiated soon after the HeMS results in a rate close to $\dot M_{\rm TH}$. 
In this case, as the mass transfer  rises above $\dot M_{\rm TH}$, a delayed dynamical 
instability develops. 

Our calculations have shown that stars less massive than 3.5 $\msun$ can 
evolve into a common envelope during the second mass transfer  event.  For these 
sequences, the two stages of mass transfer take place after core-C and core-O
ignition  with the typical orbital periods 
corresponding to about 0.2\,d at the onset of the 
second mass transfer  event.
Since, in this case, a significant amount of mass has already 
been lost during the evolution (see Table 1), it is highly 
likely that, if a CE is formed, the binary will survive the spiral-in 
phase and form a DNS.   This follows from the fact that the orbital 
energy is significantly greater (by more than a factor of 10) than 
the energy required to drive the remaining envelope matter to infinity.   As 
a result, the orbital shrinkage associated with this CE phase is not 
expected to differ very much from that predicted by the stable mass 
transfer sequences. 

A summary of our results based on the outcome of the mass transfer  phase is illustrated
in Figure 3.  We have found four qualitatively different outcomes, which 
depend on the mass of the helium star and its evolutionary stage or 
orbital period at the onset 
of mass transfer.  These outcomes can be described as: 

\begin{itemize}
\item[(i)]{ a delayed dynamical instability develops, probably leading 
to a merger (stars in Figure~3);}
\item[(ii)]{ mass transfer  proceeds stably, the mass transfer 
rate exceeds $\dot M_{\rm crit}$ during He-shell burning (hereafter
CEB case), and
probably leads to a merger (solid circles in Figure 3);} 
\item[(iii)]{mass transfer  started during or after He-shell burning and proceeds stably, but the 
mass transfer rate remains below $\dot M_{\rm crit}$  and a DNS
is formed (triangles in Figure 3);}
\item[(iv)]{ mass transfer  started during He-shell burning or after the core-C ignition
and proceeds stably, the mass transfer  rate  exceeds $\dot M_{\rm crit}$ 
after core-C ignition (hereafter
CEC case), and a DNS is
likely formed (open circles in Figure 3);}
\end{itemize}

\subsection{Population Synthesis Models }

To examine the implication of our mass transfer results for DNS
formation we have run a few indicative models. 
Based on our earlier experience with modeling double compact
object formation \cite{BKB}, we have chosen to present results for one
representative population synthesis model consisting of a population of $10^6$\  primordial
massive binary systems.
The initial conditions and binary-evolution parameters (other than He-star
mass transfer episodes) are consistent with the model A from  Belczynski et al. (2002b).
We obtain  results for 3 different sets of assumptions regarding the behavior of He stars:

\begin{enumerate}
\item Roche lobe filling He stars are treated as described in \S\,2.2.
\item He stars less massive than 5\,M$_\odot$  
that have evolved away from the HeMS experience 
stable mass transfer with mass loss and orbital period 
contraction similar to that seen in  most of our mass-transfer sequences
(and in Dewi et al. 2002).
In all other cases He stars are assumed to experience a merger. 
In the stable mass transfer episodes the donors are assumed to lose their entire 
He-rich envelopes.
Since mass transfer rates are much higher than the 
Eddington limit, we assume that all of the 
transferred material is lost from the systems 
with the specific orbital angular 
momentum of the NS (consistent with our detailed mass transfer sequences). 
During the mass transfer  phase helium stars are not nuclearly evolved 
(we refer to this as model {\it MT+CEB+CEC}).
\item similar to model 2, except that systems in the CEB region (see Figure 3)
lead to mergers (we refer to this as model  {\it MT+CEC}). 
\end{enumerate}

In the synthesis calculations one can identify among DNS progenitors those 
systems with NS and He-stars that fill their Roche lobes after core-He
burning ($\sim 80\%$ of all DNS progenitors). 
The period and He-star mass distributions of this population are shown 
in Figures 4 and 5 respectively. It can be seen that the mass distribution 
is bi-modal with broad peaks near 2.8 $\msun$  and 4 $\msun$, whereas the 
period distribution is singly peaked at an orbital period of 0.1 d with 
a tail extending to periods of greater than  1 day.
The minimum helium star mass for NS formation in our population    
synthesis models of binary evolution is $\sim 2.3 \msun$, and it agrees 
very well with earlier estimates.
We note that the lower He star mass limit for NS formation in 
single star evolution is  $\sim 2.2 \msun$ (see Habets 1986) and may extend 
to as high as $\sim 2.5 \msun$ ($2.9 \msun$) for case BB (BA) mass transfer 
in binary evolution (see Dewi et al. 2002).

In Figure 6 we show the population in terms of helium star masses $M_{\rm He; tr}$ 
and orbital periods $P_{tr}$ at the onset of mass transfer. The lines separating 
the various types of outcomes are also shown (see Figure 3). 
A comparison between Figures 3 and 6 reveals that a
significant fraction  ($\sim 47\%$) of the mass-transferring 
DNS progenitors can probably avoid a merger: 
(i) systems between solid lines (excluding dashed zones)
never develop a dynamical instability nor do they 
exceed $\dot M_{\rm crit}$ (given in eq. 15), (ii) systems within
the long-dashed line exceed $\dot M_{\rm crit}$ late enough that we 
expect them to avoid a merger.  
We have also examined two other models for 2 different values of the
$\alpha_{\rm CE}\lambda$ parameter. The fraction of the mass-transferring 
DNS progenitors that can probably avoid a merger 
are $\sim 47\%$ (for $\alpha_{\rm CE}\lambda=0.3$) and
$\sim 65\%$ (for $\alpha_{\rm CE}\lambda=3$).

In Figure 7 we present the derived distributions of DNS lifetimes (i.e., the 
time within which the two NS merge due to gravitational-wave radiation) under 
the 3 different assumptions described above.
It is evident that the two more realistic models 2 and 3 lead to 
formation of DNS with longer lifetimes than model 1.
However, the differences between the distributions do not have drastic implications.
The stable mass transfer model 2 results in about one order 
of magnitude longer merger lifetimes of DNS.
The distributions are not greatly dissimilar because the orbital contraction in the 
assumed CE evolution scenario 1 is not very dramatic due
to the small masses of the He-star envelopes (ejection of the small
envelope does not lead to drastic orbital contraction). In the 
stable mass transfer sequences (model 2) 
there is significant orbital contraction, 
because mass is transferred from the more massive to the less massive binary 
component. Consequently, short-lived DNS systems  $\sim 10$ Myr (compared to 1-10Gyr previously expected)
can still form as found in Belczynski et al.  (2002b). 
In model 3 many more DNS progenitors merge and therefore the number 
of short-period DNS is depleted compared 
to the case with stable mass transfer . 
We find DNS Galactic merger rate in 3 models considered as follows:
53, 56 and 33 DNS mergers per Myr respectively for model 1, 2 and 3.

The merger time distributions of DNS systems for both alternative CE models
with $\alpha_{\rm CE}\lambda=0$.3 and $\alpha_{\rm CE}\lambda$=3.0 are very similar.
Both distributions peak at around 10-100 Myrs with the majority of systems
having merger times between $0.1-10^{4}$ Myrs.
The distribution for $\alpha_{\rm CE}\lambda=0.3$ shows a slight overabundance of
systems with very short merger times.
This may be ascribed to the fact that all DNS progenitors have experienced CE,
and in the model with small CE efficiency, they have formed tighter systems
(and thus with shorter merger times) than for models with large CE efficiency.
In general the two additional models resemble very closely the distribution obtained  
for our standard model with  $\alpha_{\rm CE}\lambda=1.0$ shown with in Fig.7 (solid line). 
The lack of strong dependence of the merger time distribution on CE 
efficiency parameters reflects the fact that the last MT episode
leading to the formation of DNS does not involve a CE.
Therefore, by varying the values of CE parameters, we substantially alter
the population of binaries entering the last MT episode. Nevertheless this   
final MT stage acts as a filter, allowing only systems with similar
properties to eventually become coalescing DNS.

\section{CONCLUSIONS}

We have undertaken a comprehensive study  of the DNS formation via the
helium star - NS binary formation channel.  We have calculated a
suite of 60 evolutionary sequences, and delineated 
the mass $M_{\rm He; tr}$ and  orbital period $P_{\rm tr}$
regimes  where systems evolve  stably  and where  they undergo  CE
evolution.
The numerical  results reveal the classification of binary systems
depending on the type of the mass transfer  
during their evolution and the outcome of this mass transfer phase.
The classification can be described as:

\begin{enumerate}
	
\item{Unstable mass transfer that leads to  a delayed  
	dynamical instability (DDI) and a binary
	merger,  aborting DNS  formation:
	\begin{enumerate}
	\item{ Binaries with mass ratios	 
	(at the onset of the  mass transfer) greater than 3.5}
	\item{Systems with HeMS  donors (corresponding  	
	to orbital  periods less than about 0.1d
	for $M_{\rm He} < 5  \msun$ and less than 
	0.5d for $M_{\rm He} > 5  \msun$)}
	\end{enumerate}
}

\item{Stable mass transfer (possibly in 2 phases interrupted 
	by a detached phase):
	\begin{enumerate}
	\item{For Roche lobe overflow during the He-shell burning phase, the
	helium star component  can lose nearly its entire  envelope before the
	core-C ignition takes  place (first contact phase).
	The mass transfer  phase can  be interrupted at the end  of the  
	He-shell burning  stage when  the binary  usually becomes detached.   
	If this  detached stage  precedes core-C  ignition, a
	second mass transfer event ensues during and after core-C burning.  }

	\item{CE phase may occur as a result of the accretion 
	trapping  radius exceeding the Roche  lobe  radius of the NS:
		\begin{enumerate}
		\item{ in systems with $P_{\rm tr}\lesssim 0.3$ days for  $M_{He; tr} < 5 \msun$ 
		the mass transfer rate exceeds $\dot M_{\rm crit}$  during  the He-shell burning; 
		these systems are likely to merge }
		\item{in systems with $P_{\rm tr}\approx 0.1 - 0.5$ days for 
		low He-star masses ($\sim 2.6 - 3.3\,\msun$)the mass transfer 
		rate exceeds $\dot M_{\rm crit}$  after core-C ignition;
		binaries survives and a DNS is likely formed.}
		\end{enumerate}
        }
	\end{enumerate}
}
\end{enumerate}

In the case of stable mass transfer without the occurrence of the CE, typically,
the  amount of envelope  mass that remains  at the  end of  the evolution
depends on the duration of the  mass transfer:  the earlier
mass transfer  occurs, the smaller  the remaining envelope  mass.  Our
results suggest that close systems consisting of low-mass helium stars
($\lesssim 3.5 \msun$) lose  a significant fraction of their envelopes
(see Table 1),  especially if they evolve through  a second phase of
mass transfer.  Such systems are one of the possible types
of binary progenitor candidates for Type
Ic supernovae  (see also Dewi et al.  2002). However the contribution 
of this channel is only about 1\% of the total Ic supernovae 
rate (0.7\% if CE phase occurred during the He shell burning).
As a result  of the mass transfer, the  systems evolving through 
stable mass transfer form a short period population
of immediate  DNS progenitors with  periods ranging from 0.03\,d 
to 1.7\,d.

We have classified binary systems as possible mergers
if CE occurred and the  binding energy of the  common envelope is  greater than the
energy released  from the  orbit.  
Such merged  models leading  to the
formation of a  black hole and a helium rich  accretion disk have been
discussed  as potential progenitor  candidates of  $\gamma$-ray bursts
(Fryer \& Woosley 1998, Zhang \& Fryer 2001).  These mergers 
may increase  previously predicted
rates  of  $\gamma$-ray  bursts   (Fryer,  Woosley  \&  Hartman  1999;
Belczynski et al.  2002c). The rate is about 0.5\% of the total
supernovae rate (all types), if one assumes that only mergers
of He stars more massive 
than 4\,M$_\odot$ produce $\gamma$-ray bursts 
(0.7\% if CE events occurred during the He shell burning). 
If one assumes that $\gamma$-ray bursts can be produced 
by NS merger with  the He star more massive than 2.3\,M$_\odot$,
the corresponding rates are 1.7\% and 2.2\% of the total supernova rate.

We  note that  the CE  phase  could be  avoided if  the critical  mass
transfer rate occurs  after the core-O ignition  when the nuclear
evolutionary timescale becomes shorter than the spiral-in timescale.

The short orbital  periods of the immediate progenitor  systems of DNS
have important implications for  the distribution of DNS merger times.
In  comparison to  Belczynski et  al. (2002b),  it is  found  that the
fraction of  very short-lived ($<1$\,Myr) DNS  systems decreases since
the majority of DNS progenitors  evolve through a stable mass transfer
phase rather than through a  CE phase with the survival.  Although the
peak of the distribution occurs  at merger times $\sim 2-12$ Myr, very
short-lived  systems  are  still  formed at  significant  rates  since
orbital contraction is non-negligible for the stable mass transferring
systems.  In particular, for the models that we examined, the fraction
of short-lived ($<1$\,Myr) DNS binaries is reduced from $\simeq 50 \%$
(27 mergers per Myr in our Galaxy)  to $\simeq 15\%$ (8 mergers per Myr in
our Galaxy).   As  pointed  out  by  Belczynski  \&  Kalogera  (2001)  and
discussed  also by  Belczynski et  al. (2002b),  a population  of such
short-lived  DNS has  implications  for the  estimates of  coalescence
rates based on  the currently observed DNS sample,  which consists only
of long-lived  systems ($>>1$\,Myr).  In view of  our results  in this
study,  the associated  upward correction  factors for  such empirical
rate estimates  are modified and are  $\simeq 1.3$ (instead  of 2.5 as
estimated  by Belczynski et  al. (2002b).   However, in the  case where CE
phase occurs during He shell  burning, the number of short-period
binaries vanishes.

The lifetimes of DNS systems  also have important implications for the
typical  galactocentric  distances  of  DNS  merger  sites  and  their
possible  association with  $\gamma$-ray bursts  (Bloom,  Kulkarni, \&
Djorgowski  2002).    In  view  of  this  reduced   fraction  of  very
short-lived DNS,  we have re-examined the  cumulative distributions of
the  projected  galactocentric  distances  of DNS  merger  sites  [for
details see  Belczynski et al. (2002a)].   In Figure 8  we present the
distributions for the 3 cases considered in \S\,3.2.   
The initial  distribution of  primordial binaries  in a
galaxy  is also  shown  for  comparison.  The  farther  a given  model
deviates from the initial distribution, the farther is the location of
merger sites from the center of the galaxy.  Calculations for galaxies
similar in size and mass to the Milky Way, and for dwarf galaxies with
$1\%$  of the Milky Way mass have  been performed [for details
see Belczynski et al. (2002c)].

It is evident that, in the  case of large galaxies like the Milky Way,
there is  little difference compared  to the results of  Belczynski et
al. (2002a) and  the DNS mergers, whichever model  is used, take place
inside  the larger,  massive  hosts.  Even  for  moderately long-lived
systems, the deep gravitational  potential of massive galaxies as well
as  its large  size prevents  DNS from  escaping far  from  the galaxy
borders  before  their  coalescence  due  to  gravitational  wave
radiation.  Therefore, even for two different populations of DNS (with
very short  merger times  for CE evolutionary  model and  short merger
times  for  stable mass transfer   model)  their  merger  locations are  basically
indistinguishable.

For  the  smallest and  lowest  mass  galaxies,  there are  noticeable
differences between  the distributions.   Since low-mass galaxies
have  a  very  small  effect  on  fast  moving  DNS,  even  small
differences in  merger times are  imprinted on the location  of merger
sites.  The longer the merger  time the farther the system will escape
from the  host galaxy.   For the model in which the
immediate DNS progenitor survive CE evolution, model 1, 
only a small fraction ($\sim 10 \%$) of DNS
will merge outside  the low mass hosts. On the  other hand, for models
representing our  new calculations we find  that up to $20  \%$ of DNS
mergers may take  place outside such galaxies ($30 \%$  in model 2).  
Nevertheless, this is in stark
contrast to the previous population synthesis results, predicting that
as many  as $\sim 50-80 \%$  of DNS should escape  their host galaxies
(e.g.,  Bloom et al.   1999; Portegies  Zwart et  al.  1999;  Bulik et
al. 1999).   Therefore, assuming that some 
$\gamma$-ray bursts  are connected to DNS mergers, it is most 
likely that they occur within their galaxies.

\acknowledgments
 We would like to thank C.~Fryer, N.~Langer, O.~Pols, 
and R.F.~Webbink for useful discussions. 
This work is partially supported by the Lindheimer Fund at Northwestern
University, and NSF grant PHY-0121420 to VK, NSF grant PHY-0133425 to FR, 
and AST-9727875 and AST-0200876 to RT.
VK also acknowledges support by the David and Lucide Packard Foundation
through a Science and Engineering Fellowship.

\newpage

\begin{deluxetable}{l l l l l l l l l l l l }
\tabletypesize{\scriptsize}
\tablecaption{
The binary parameters and outcomes of representative model
sequences.  The columns denote: the mass, $M_{\rm He;0}$\ (in $\msun$) 
on the zero age helium 
main sequence, the mass, $M_{\rm He; tr}$ ( in $\msun$),
the mass of the envelope, $M_{\rm env; tr}$ ( in $\msun$),
 orbital separation,
 $A_{\rm tr}\ ({\rm in\ } R_\odot)$, 
orbital period  $P_{\rm tr}$ (in days), 
at the onset of mass transfer;  $\dot M_{\rm TH}$ (in $M_{\odot} {\rm yr}^{-1}$) is 
the thermal timescale mass transfer rate. The evolutionary
stages are denoted as case 
A for Roche lobe overflow on HeMS, case B during  He-shell burning, and 
case C for phases after carbon ignition.  
The possible outcomes are: 
DDI for delayed dynamical instability (that leads to a merger), 
CEB for possible CE formation during He-shell  burning 
(likely to lead to a merger), 
CEC for possible CE formation after core-C ignition (DNS is likely formed), 
MT  when CE did not occur (DNS is formed), 
and NC denoted the boundary case 
when the system did not interact (for all periods larger 
than cited in the table, mass transfer  will not occur). 
For models that do not experience DDI, but interact, we list the 
final period, $P_{\rm f}$ (in days), 
the final mass of the helium star, $M_{\rm He;f}$\ (in $\msun$), 
the final mass of the envelope, $M_{\rm env;f}$ (in $\msun$), 
and the duration of the mass transfer  episode, $\Delta t_{\rm tr}$ 
in years. The duration of mass transfer  is 
calculated from the onset of Roche lobe overflow to the ignition of oxygen 
in the core (end of model calculations) or to the detachment of the He star from 
its Roche lobe (MT cases), whichever occurs earlier.
If only one mass transfer  phase was encountered, there is one entry in the 
$\Delta t_{\rm tr}$
column. A second entry indicates that the mass transfer  stopped but was 
then reinitiated 
and the duration of the second mass transfer  phase is given.
The first entry in the $M_{\rm env;f}$ column always represents its value at the end
of the first mass transfer  phase.  The second 
entry, if present, indicates that the first mass transfer  phase finished before the
end of our evolutionary calculations (core-O ignition). 
We list the envelope mass at the end of the model calculations, depleted (in
relation to the first entry) either through the second mass transfer  event or by 
wind mass loss. 
}
\tablehead{
\colhead{$M_{\rm He;0} $}      & \colhead{$M_{\rm He; tr} $} &
\colhead{$M_{\rm env; tr} $} & \colhead{$A_{\rm tr} $} &
\colhead{$P_{\rm tr} $}      & \colhead{$\dot M_{\rm TH} $}     & 
\colhead{Case} &  \colhead{outcome} & \colhead{$P_{\rm f} $}        
& \colhead{$M_{\rm He; f} $}   & \colhead{$M_{\rm env; f} $}  
& \colhead{$\Delta t_{\rm tr} $}   }
\startdata
6.0 & 4.99 & 1.91 & 0.99 & 0.045& 1.7$\cdot 10^{-4}$ & B  & DDI & -- & -- & -- & --  \\*
6.0 & 4.97 & 1.88 & 1.68 & 0.1 & 4.0$\cdot 10^{-4}$ & B  & DDI & -- & -- & -- & --  \\*
6.0 & 4.96 & 1.85 & 2.66 & 0.2 & 6.8$\cdot 10^{-4}$ & B  & DDI & -- & -- & -- & --  \\*
6.0 & 4.95 & 1.80 & 3.43 & 0.3 & 9.2$\cdot 10^{-4}$ & C & CEC & 0.23 & 3.94 & 0.69 & 2.3$\cdot10^{3}$ \\*
6.0 & 4.95  & 1.78 & 4.15 & 0.4 &  1.1$\cdot 10^{-3}$ & C  & DDI & -- & -- & -- & --  \\*
6.0 & 4.95  & 1.68 & 5.0  & 0.52 & -- & --  & NC & -- & --  & -- & -- \\
\\ 
5.0 & 4.32 & 1.99& 1.45  & 0.085 & 3.2$\cdot 10^{-4}$&B  & DDI & -- & -- & -- & --  \\*
5.0 & 4.32 & 1.82 & 1.6  & 0.1 &3.2$\cdot 10^{-4}$ &B  & CEB & 0.027 & 2.88 & 0.57/0.3& 3.3$\cdot10^{3}/34$\\*
5.0 & 4.32 & 1.80 & 3.35  & 0.3 &8.2$\cdot 10^{-4}$  &B  & CEB & 0.088 & 3.01 &0.73/0.42& 5.1$\cdot10^{3}$ \\*
5.0 & 4.31 & 1.76 & 4.00  & 0.39 &1.0$\cdot 10^{-3}$ &C  & MT & 0.14 & 3.24 & 0.58 & 5.5$\cdot10^{3}$\\*
5.0 & 4.29 & 1.67 & 9.2  & 1.4 & -- &-- & NC & -- & --  & -- & -- \\
\\
4.0 & 3.62 & --    & 1.3  & 0.075 &  --  &A  & DDI & -- & -- & -- & --  \\*
4.0 &3.58 & 1.60& 1.46 & 0.1 &2.5$\cdot 10^{-4}$ &B & CEB & 0.041 & 2.34 &0.73/0.3 & $4.8\cdot10^{3}/1\cdot10^{4}$\\*
4.0 &3.54 & 1.53 & 2.31 & 0.2&4.6$\cdot 10^{-4}$ &B & CEB & 0.089 & 2.49 & 0.49/0.38  &$1.1\cdot10^{4}$\\*
4.0 &3.51 & 1.50 & 3.05 &0.3&6.4$\cdot 10^{-4}$ &B & MT & 0.14 & 2.52 &0.58/0.38 & $7.8\cdot10^{3}/1.3\cdot10^{3}$\\*
4.0 & 3.50&1.46 & 3.76& 0.4 &8.2$\cdot 10^{-4}$ &B  & MT & 0.21 & 2.62 &0.60/0.45&$6.5\cdot10^{3}/1.3\cdot10^{3}$ \\*
4.0 & 3.50 &1.46 & 4.96  & 0.6 &1.1$\cdot 10^{-3}$ &C  & MT & 0.16 & 3.46 & 0.51& $1.1\cdot10^{4}$\\*
4.0 & 3.50 & 1.46 & 7.0  & 1.0 &1.5$\cdot 10^{-3}$ &C  & MT & 0.98 & 3.47 & 1.42 & 520\\*
4.0 & 3.50 & 1.46 & 17.0  & 3.7 & -- &--  & NC & -- & -- & -- & --  \\
\\ \ 
\\ \ 
\\ \ 
\\ \ 
\\ \ 
\\ \ 
\\ \ 
\\ \ 
\\ \ 
\\ \ 
\\ \ 
\\ \ 
\\ \ 
\\ \ 
\\ \ 
\\ \ 
\\ \ 
\\ \ 
\\ \ 
\\ \ 
\\ \ 
\\ \ 
\\ \ 
\\ \ 
\\ \ 
\\ \ 
\\ \ 
\\ 
3.5$$ &  3.22 & -- & 1.07  & 0.06 & -- &  A & DDI & -- & -- & -- & -- \\*
3.5 &3.21 & 1.55&1.49& 0.1 &2.2$\cdot 10^{-4}$ & B & MT & 0.055& 1.94 &1.08/0.12 &$7.8\cdot10^{3}/2.4\cdot10^{4}$\\*
3.5 &3.19 &1.49&2.37& 0.2 &4.4$\cdot 10^{-4}$ & B & MT & 0.12 & 2.11&0.47/0.14&$1.9\cdot10^{4}/2.9\cdot10^{3}$\\*
3.5 &3.19 & 1.48&3.12& 0.3 & 5.1$\cdot 10^{-4}$& C &CEC & 0.17 & 2.08 & 0.50/0.21&$1.2\cdot10^{4}/8.1\cdot10^{3}$\\* 
3.5 &3.19 & 1.48 &3.75& 0.4 &7.6$\cdot 10^{-4}$ & C & CEC & 0.23 & 2.13 &0.54/0.25&$8.3\cdot10^{4}/7.8\cdot10^{3}$\\* 
3.5 &  3.19 & 1.48 &4.84  & 0.6 &1.0$\cdot 10^{-3}$ & C & MT & 0.23 & 2.177 & 0.32 & $1.9\cdot10^{4}$\\*
3.5 &  3.16 & 1.27&27.8  & 8 & -- & -- & NC & --& -- & -- & -- \\ 
\\ 
3.0 &  2.91 & -- & 1.06 & 0.06  & -- &A  & DDI & --   & -- & -- & --  \\*
3.0 &  2.80 & 1.39 & 1.50 & 0.1  &1.9$\cdot 10^{-4}$ &B  & CEC & 0.08&1.69&0.30/0.10&$5.7\cdot10^{4}/770$  \\*
3.0 &2.78 & 1.37& 2.37 & 0.21  &3.8$\cdot 10^{-4}$& B&CEC & 0.16 & 1.76&0.38/0.11&$2.8\cdot10^{4}/2.5\cdot10^{3}$\\*
3.0 &  2.78 & 1.36& 2.98 & 0.3  &5.3$\cdot 10^{-4}$ &B& MT & 0.22& 1.86&0.42/0.22&$1.8\cdot10^{4}/5.6\cdot10^{3}$\\*
3.0 &  2.77 &1.32 & 3.66 & 0.4  &6.4$\cdot 10^{-4}$& B &MT&0.3&1.81&0.46/0.16&$1.3\cdot10^{4}/1.1\cdot10^{4}$  \\*
3.0 &  2.77 & 1.22& 10.6 & 2.0  &2.0$\cdot 10^{-3}$ &C  & MT & 1.7   & 2.54 &0.96& $2.6\cdot10^{3}$\\*
3.0 &  2.76 & 1.14 & 35.8 & 25.0  & --& --  & NC & --   & -- & -- & --  \\ 
\enddata
\label{table}
\end{deluxetable}

\begin{figure} 
\plotone{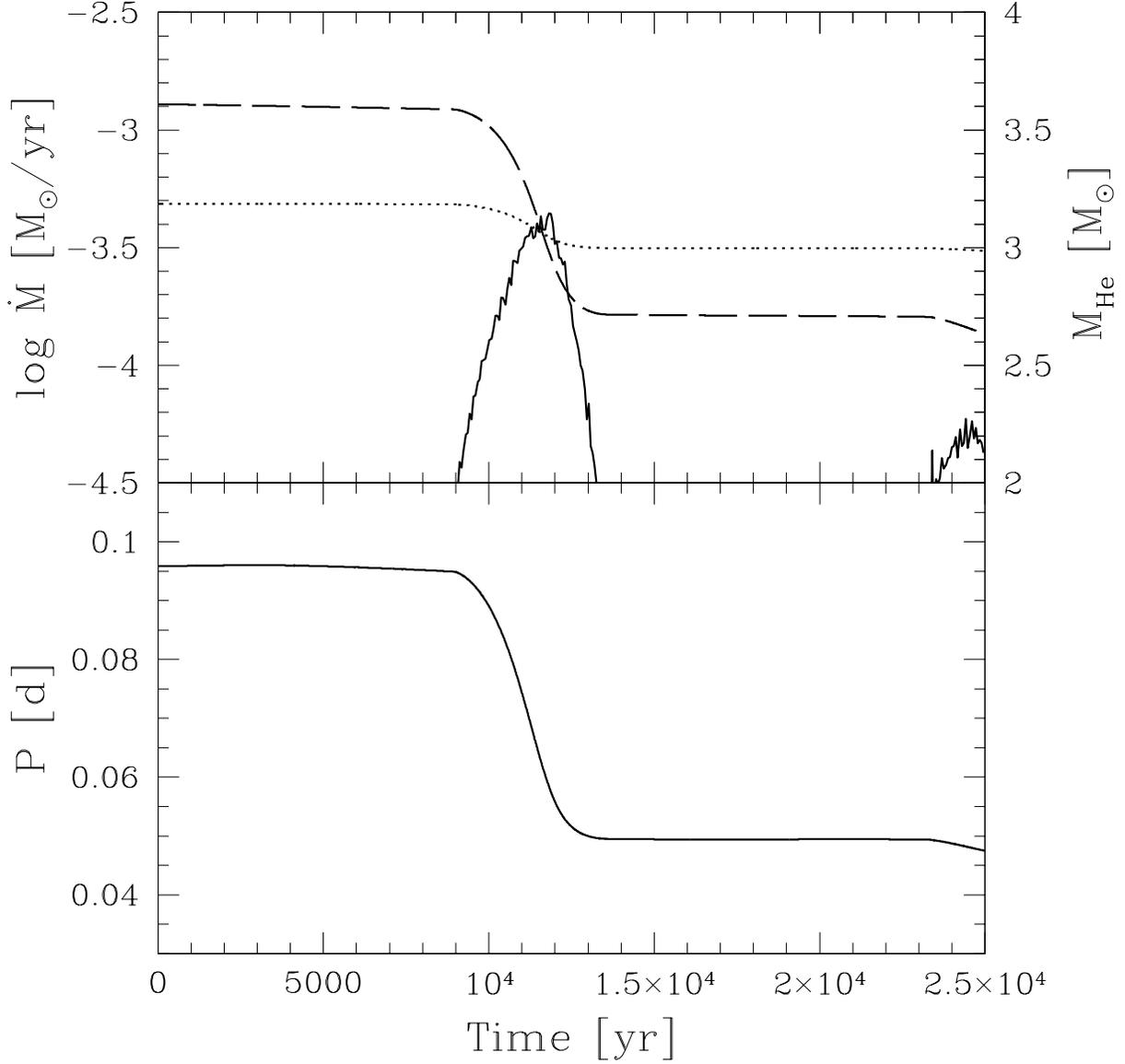} 
 \caption{Evolution of a $4 \msun$ He-star 
(mass at the zero age helium main sequence) and $1.4 \msun$ NS binary 
with an orbital period at the onset of mass transfer  of 0.096 d. 
Mass transfer begins during 
He-shell burning for the He-star donor.
The time evolution of the mass transfer  rate and the 
He-star mass are shown in 
the upper panel (solid and dashed lines, respectively) and the binary 
orbital period evolution is presented in the lower panel. 
The critical mass transfer  rate ({eq.\ref{mcrit}}) for possible onset of a CE 
phase is also shown in the upper panel (dotted line).
Note also the second mass transfer phase initiated after core-C ignition 
(lower right corner in the upper panel).
The zero of time corresponds to the start of the mass transfer
sequence when the age of the primary is 1.4$\cdot 10^6$ years}
\end{figure}

\newpage

\begin{figure}
\plotone{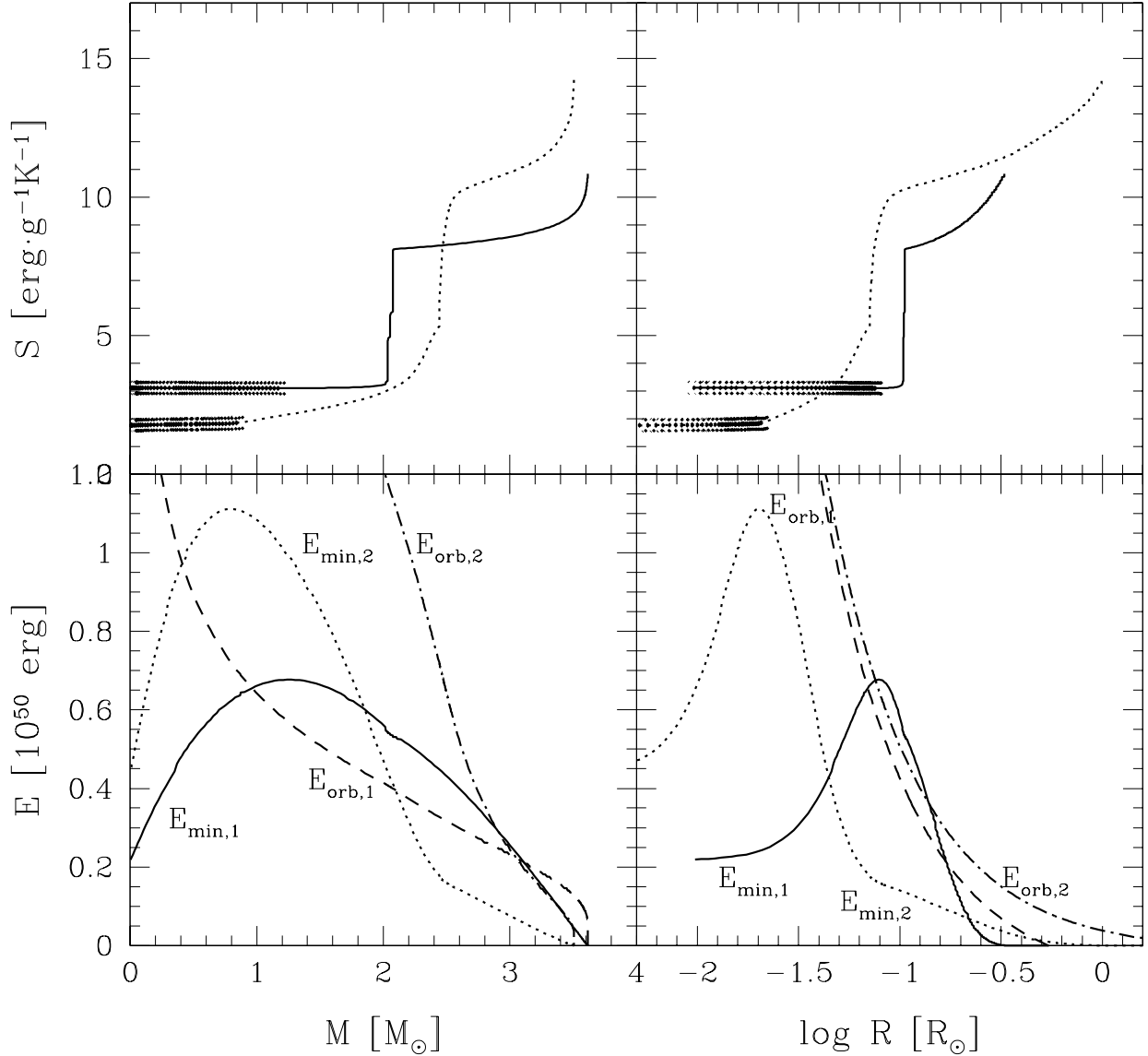}
 \caption{ Specific entropy (upper panels) and energies (lower panels)
as functions of mass (left panels) and radius (right panels)
for a 4\,M$_\odot$ He-star at two different 
evolutionary stages: soon after the HeMS (solid lines in 
upper panels, index 1 in lower panels) 
and during core-C burning (dotted lines in upper panels, index 
2 in lower panels).
Hatched regions show convective zones; note that convection develops only
in the He-star core. 
Note also that the CE survival is likely only for the evolved He-star
(core-C burning), since the orbital energy (E$_{\rm orb}$) is significantly 
greater than the energy needed to unbind its envelope (E$_{\rm min}$).
Orbital energies were calculated for a Roche-lobe filling 
4\,M$_\odot$ He-star (initial mass) in orbit around
a 1.4\,M$_\odot$ NS. The corresponding orbital periods are
0.075 and 1.2 days  for models 1 and 2, respectively. }
\end{figure}

\newpage

\begin{figure}
\plotone{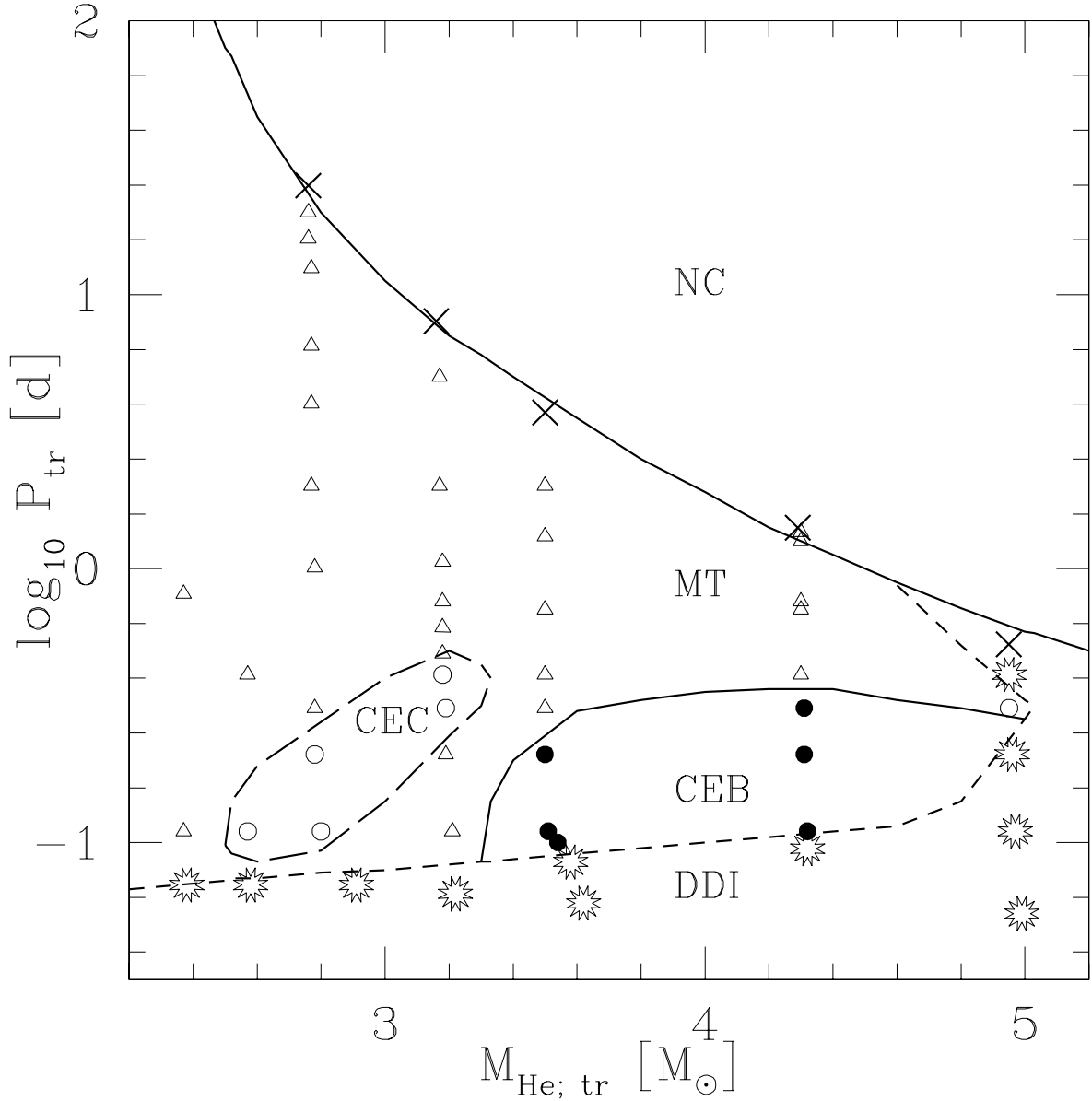} 
 \caption{He-star masses $M_{\rm He; tr}$ and orbital periods $P_{\rm tr}$ 
at the onset
of mass transfer. Different symbols correspond to different outcomes of
the mass transfer sequences:  {\em crosses} -- mass transfer never initiated;  
{\em triangles} -- mass transfer rate never exceeds $\dot M_{\rm crit}$ and
a DNS is formed; {\em
open circles} -- mass transfer rate exceeds $\dot M_{\rm crit}$ after core-C
ignition (case C) and a DNS is likely formed; 
{\em filled circles} -- mass transfer rate exceeds $\dot
M_{\rm crit}$ during He-shell burning (case B) and the
binary system will likely merge; 
{\em stars} -- mass transfer leads
to delayed dynamical instability and the binary system will likely merge. 
Lines delineate the boundaries between
different outcomes in the $M_{\rm He;tr} - P_{\rm tr}$ parameter space. }
 \end{figure}

\newpage

\begin{figure}
\plotone{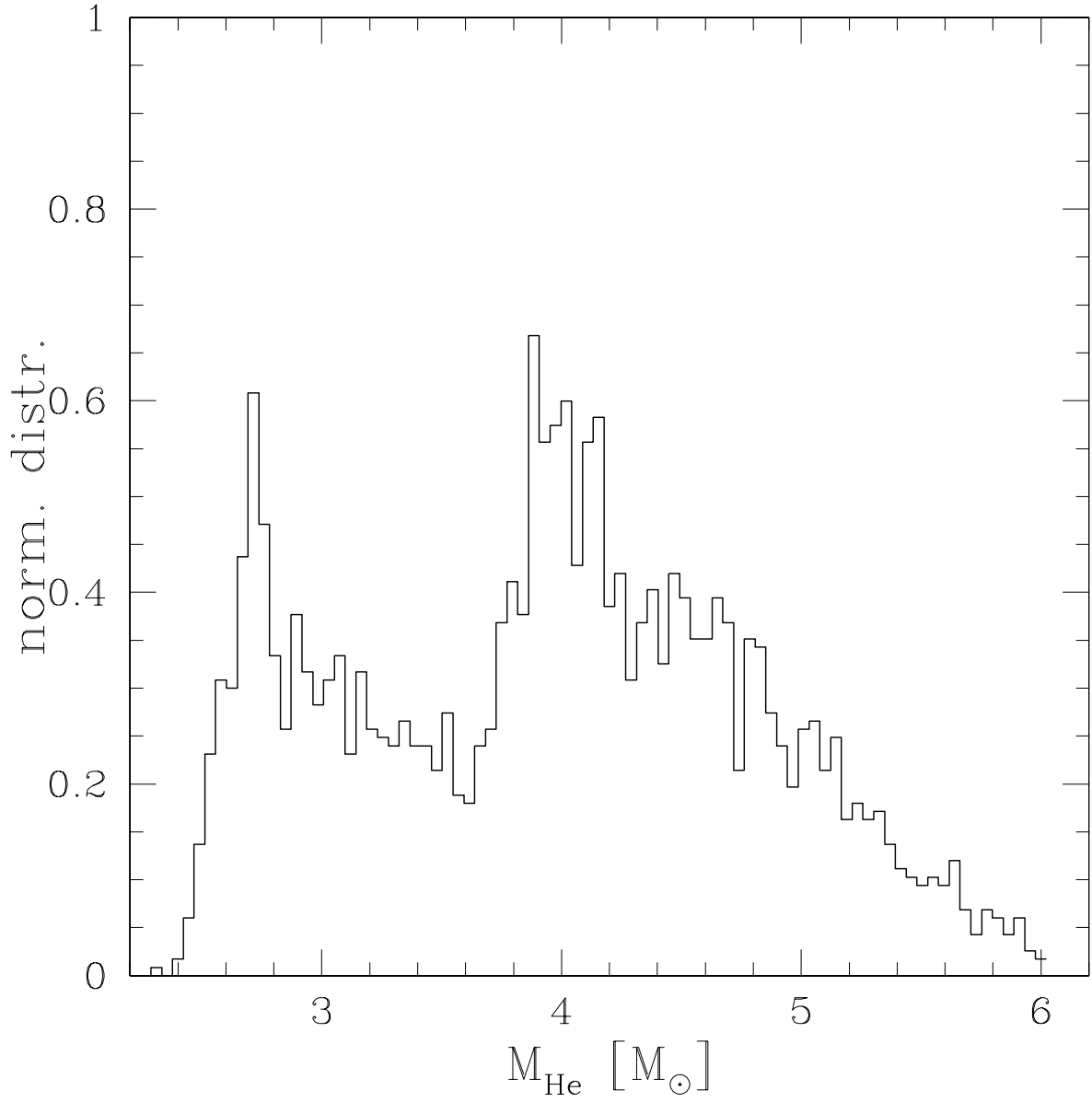} 
 \caption{Normalized distribution of He-star masses at the onset 
of the mass transfer  episode in potential DNS progenitors.}
\end{figure}

\newpage

\begin{figure}
\plotone{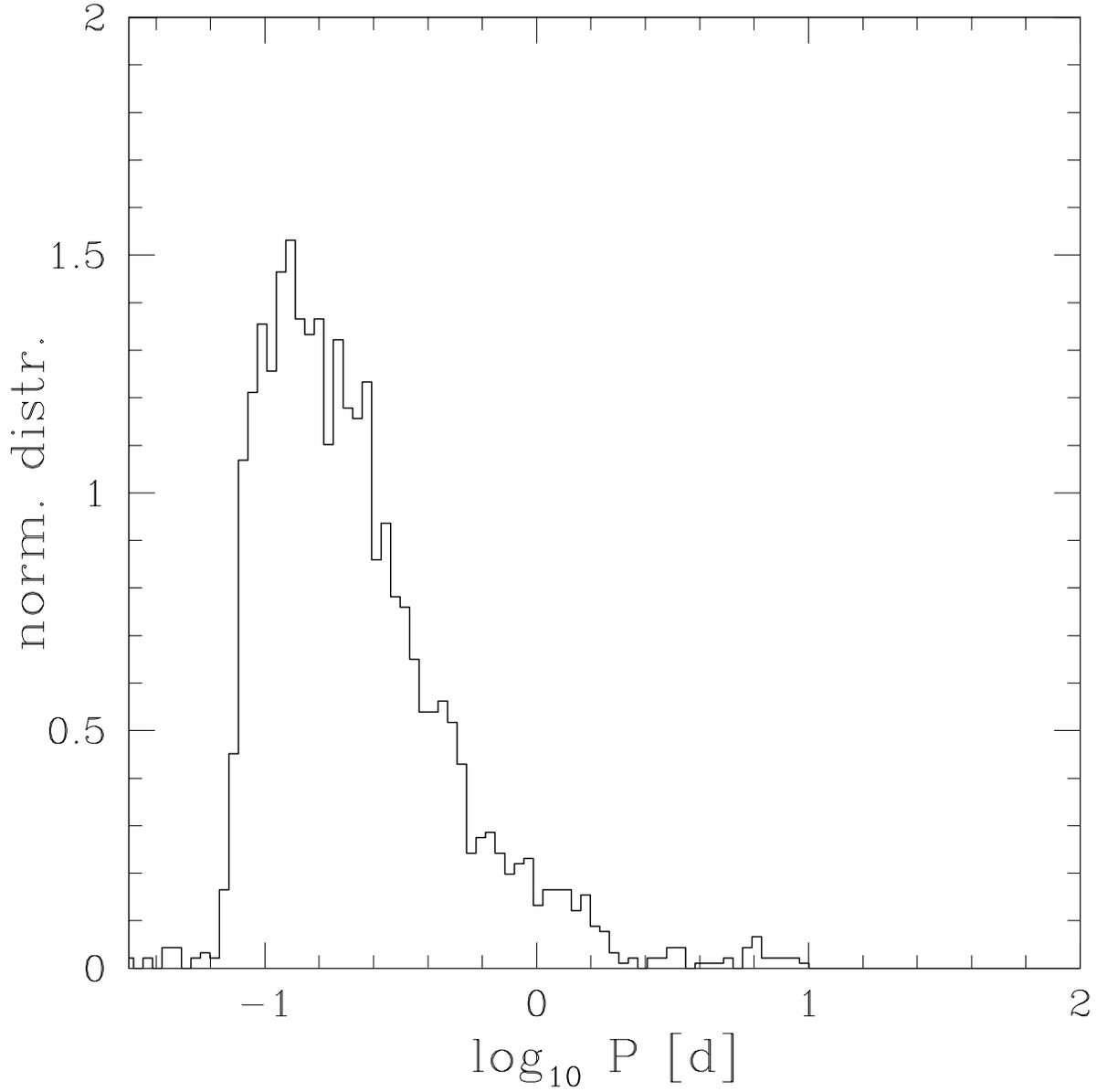}
 \caption{Normalized distribution of of 
orbital periods  of potential 
DNS progenitors at the onset 
 of the mass transfer episode.}
\end{figure}

\newpage

\begin{figure}
\plotone{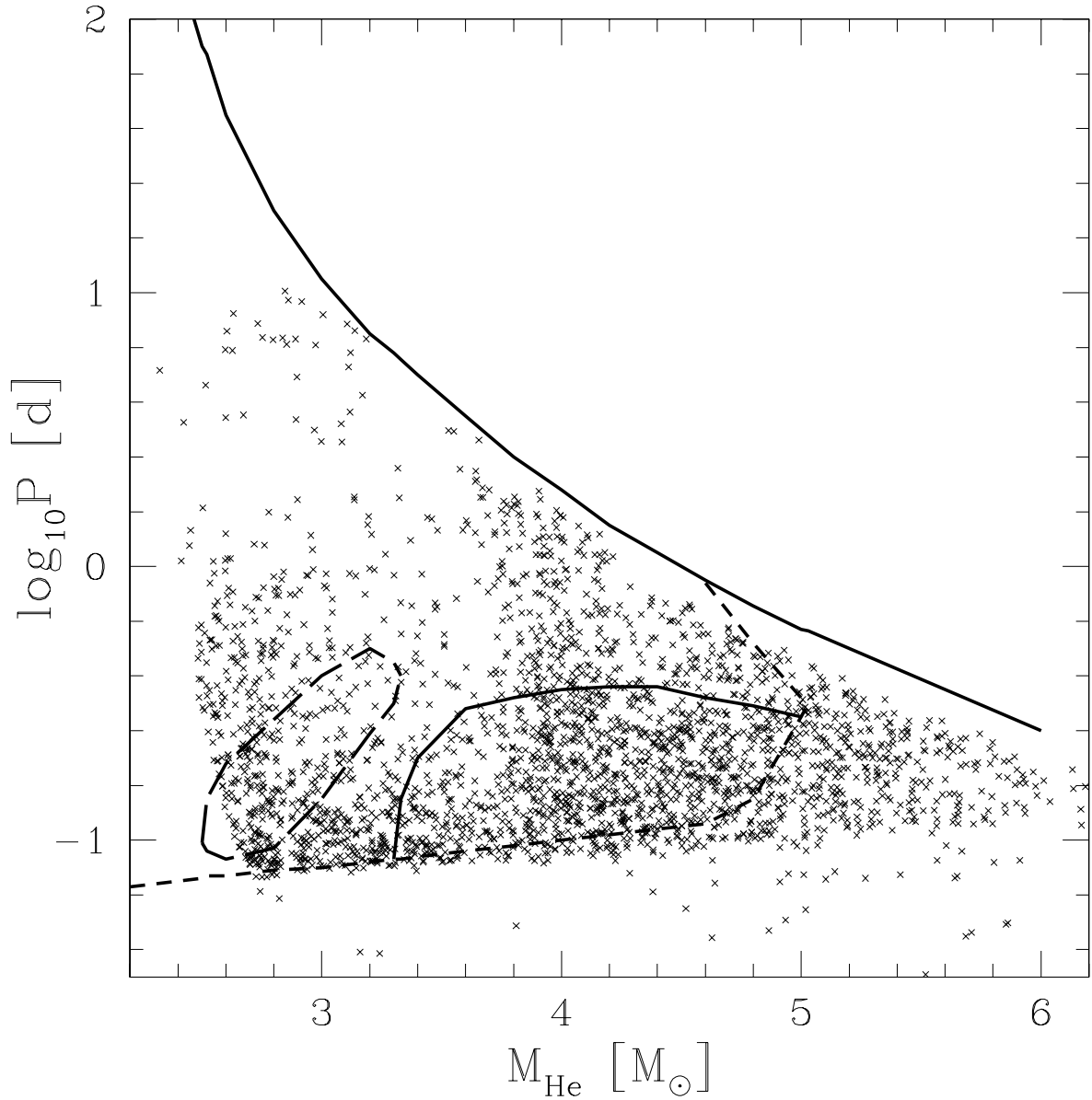}
\caption{He-star masses and orbital periods of potential 
DNS progenitors that reach 
Roche lobe overflow after the end of core-He burning (HeMS),
at the onset of the mass transfer episode. 
Lines correspond to the boundaries in Figure~3.   } 
\end{figure}

\newpage

\begin{figure}
\plotone{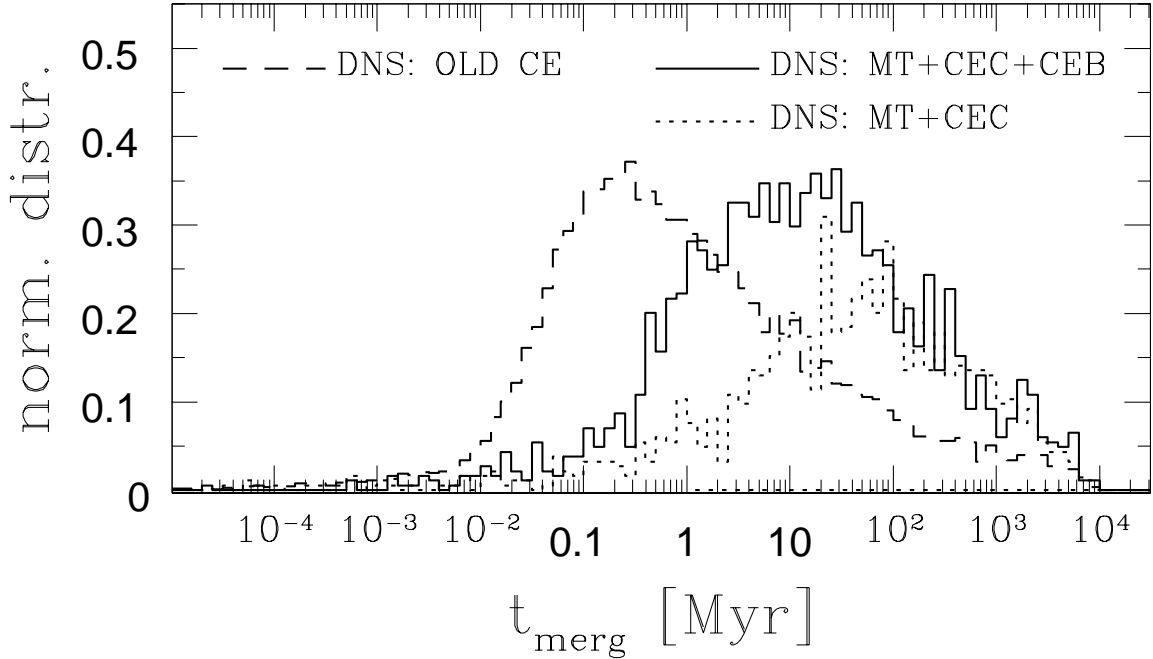} 
\caption{Normalized distribution of DNS merger lifetimes (due to
gravitational radiation) under three different sets of 
assumptions about the outcome
of mass transfer for He stars in binaries (from Figure~4): (i) systems with
evolved He-star donors with $M_{\rm He} \leq 4.5 M_\odot$\ go
through CE evolution (as assumed in  Belczynski et al. 2002b; dashed line), 
(ii) systems in the regions marked as MT, CEC and CEB 
in Figure~3 go through  stable (but 
highly non-conservative) mass transfer (solid line), 
(iii) systems  in the regions marked as MT and CEC go through  stable 
mass transfer (dotted line), whereas systems from the region
marked as CEB are assumed to merge.}
\end{figure}

\newpage

\begin{figure}
\plotone{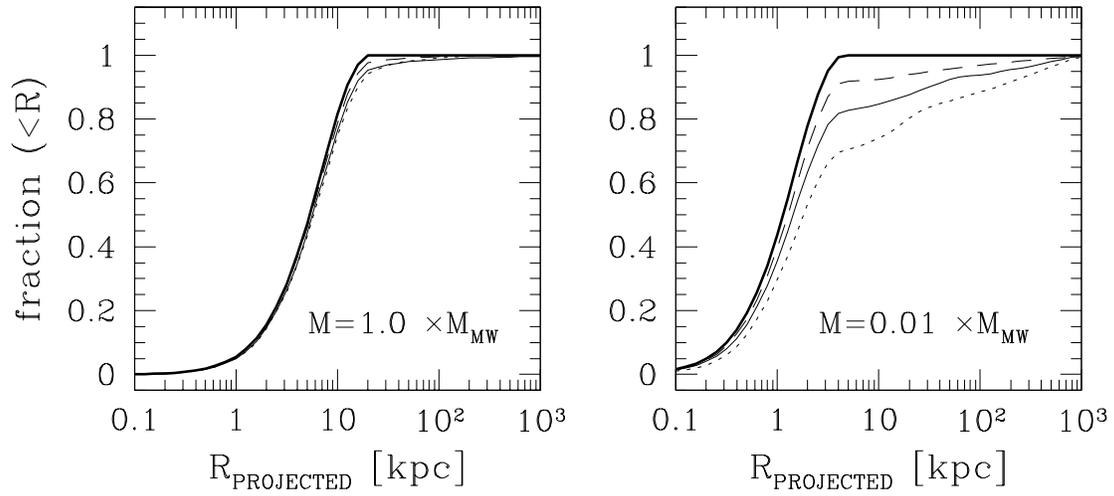}
\caption{Cumulative distributions of DNS merger sites
around a massive galaxy with $M=M_{\rm MW}=1.5\cdot10^{11}\,M_\odot$ 
(left panel)   
and a dwarf galaxy with $M=0.001\times M_{\rm MW}$ (right panel).
Thick solid lines correspond to the initial galactic distribution of
binaries. 
Dashed lines correspond to DNS formed through CE evolution (assumed in 
Belczynski et al. 2002b), while thin solid and dotted lines  correspond to
DNS formed through stable mass transfer from the MT, CEC, and CEB groups
or from only from the MT and CEC groups as marked in Figure~3, respectively.}
\end{figure}

\end{document}